# Multi-Layer Spectral Clustering Approach to Intentional Islanding In Bulk Power Systems

Faycal Znidi, Hamzeh Davarikia, Kamran Iqbal, Masoud Barati

**Abstract** Intentional controlled islanding (ICI) is a final resort for preventing a cascading failure and catastrophic power system blackouts. This paper proposes a controlled islanding algorithm that uses spectral clustering over multi-layer graphs to find a suitable islanding solution. The multi-criteria objective function used in this controlled islanding algorithm involves the correlation coefficients between bus frequency components and minimal active and reactive power flow disruption. Similar to the previous studies, the algorithm is applied in two stages. In the first stage, groups of coherent buses are identified with the help of modularity clustering using correlation coefficients between bus frequency components. In the second stage, the ICI solution with minimum active and reactive power flow disruption and satisfying bus coherency is determined by grouping all nodes using spectral clustering on the multi-layer graph. Simulation studies on the IEEE 39-bus test system demonstrate the effectiveness of the method in determining an islanding solution in real time while addressing the generator coherency problem.

## 1 Introduction

Intentional controlled islanding (ICI) has been proposed as a corrective measure of last resort to split the power system into several sustainable islands and prevent cascading outages. Most approaches to islanding aim to find, as a primary objective, electromechanically stable islands with minimal load shedding. To find a reasonably good islanding solution, all subsystems must satisfy some constraints, such as power flow disruption, generator coherency, transient stability, etc. [1].

Traditionally, the islanding problem has been solved using combinatorial optimization approaches. The inclusion of reactive power or voltage in the constraints leads to a mixed integer nonlinear program (MINLP), which is, in general, difficult to be solved than nonlinear programming problem (NPP) and mixed integer linear programming (MILP) problem [1, 2]. Therefore, linear DC power flow has often been used in literature resulting in a MILP problem that promises a better computational burden. Additionally, some other methods consider only the active power in system partitioning. As an example, in [2], it is claimed that local reactive power compensators can be used to compensate reactive power imbalance. In [3], a MILP-based splitting strategy is proposed to manage energy production and demand. In this methodology the reactive power is viewed as a local issue and can be handled with local reactive power compensators and only active power is considered in the splitting scheme. However, the reactive power plays a significant role in supporting the voltage profile, and a significant mismatch of the reactive power supply and demand causes high or low voltage conditions within islands. In [4], the MILP-based optimization method for controlled islanding disregards the generator coherency constraint, which is one of the most important requirements in the islanding solution. The optimization-based islanding algorithms are proposed in [5, 6], aiming to find the boundaries of electric islands. Utilizing the mathematical programming for islanding solution requires different set of constraints to ensure the islands integrity and feasibility, including not limited to power balanced, connectivity, and operational constraints. In the other hand, the graph-based islanding solutions automatically satisfies the connectivity constraints, since the solution is sought trough the minimum cuts in the graph.

In [7], the constrained spectral clustering is used to find islanding boundary with minimal power flow disruption. In [8], a binary particle swarm optimization (BPSO) seeking Pareto non-dominated solutions algorithm is presented to find islands containing coherent generator groups with minimal power imbalances. BPSO being a stochastic evolutionary algorithm, multiple runs of the algorithm are needed to determine if the results are consistent. In [9], a two-step spectral clustering controlled islanding algorithm is introduced, while using generator coherency as the sole constraint with minimum active power flow disruption objective to find a suitable ICI solution. In [10, 11], the authors presented an islanding scheme with minimal active power flow disruption using a constrained spectral embedded clustering technique, while satisfying the generator coherency constraints. However, these techniques disregard the effects of the bus voltage magnitude, and reactive power, which has a substantial impact on the dynamic coupling. In [12], a methodology based on dynamic frequency deviations of both generator and non-generator buses, with respect to the system nominal frequency is presented. Overall the center of inertia concept has shown its advantages in various applications.

While there is an in-depth treatment of individual topics such as generator coherency, optimization, and the active and reactive power graph-based models in the literature, there is a dearth of information regarding these multiple topics in a single model [13-17].

In this paper, a multi-layer graph spectral clustering controlled islanding (M-SCCI) algorithm for ICI solution of power systems is presented. In the first stage of the algorithm, the frequency similarity of buses, the active power flow between buses, and the reactive power flow between buses construct three different layers of the multi-layer graph. The frequency similarity among each pair of buses is evaluated using correlation among the bus frequency components. To determine the number of islands, the modularity clustering is applied to the layer containing frequency similarities among buses, which results "$k$" numbers of coherent buses or coherent groups of generators. The number of "$k$" cluster outcomes of this grouping serves as the input in the second stage of the M-SCCI algorithm that identifies island boundaries with minimal active and reactive power flow disruption. This technique is based on a multi-layer graph, whose common vertex set represents the buses, and the edges on individual layers represent power system attributes that reflect the similarities among the buses in term of the various modalities. These modalities include: ① frequency correlation coefficient between buses; ② real power flow disruption; ③ reactive power flow disruption.

## 2 Graph theory approach to controlled islanding problem

### 2.1 Multi-layer graph models of power systems

An electrical network is undirected graph $G(V, E, w)$ where each element $v_i \in V$ is either a substation or a transformer, the edge $e_{i,j} = (v_i, v_j) \in E$ is a physical cable between two nodes [18, 19], and $w$ is the associated edge weight. A multi-layer graph $G$ consists of $M$ distinct graph layers $G_i, i = 1,2,\dots,M$, where each distinct layer $G_i = \{V, E_i, w_i\}$ is a undirected and weighted graph composed over a common vertex set $V$ and particular edge set $E_i$ with associated weights $w_i$ [20]. The sets comprising the graph assume interest from an operational and physical point of view. The individual layers characterize specific relationships among entities, such as the frequency similarity associated with each pair of the island's buses, and the active and reactive power flow disruption.

A generic representation of the three-layer graph in power system is depicted in Fig. 1, where the first, second and third layers are associated to the frequency similarity, active power flow, and reactive power flow, respectively. These layers have the same nodes which represent the buses in power networks, while the edges are associated to the frequency similarity between buses in layer one, the active power flow between buses in layer 2, and the reactive power flow between buses in layer three. While the first layer is a full weighted graph (all the nodes are connected with each other), the other layers have the same edges as physical lines in power networks.

### 2.2 Dynamic generator coherency

Following a sudden disturbance on the power grid, the dynamic response of individual generators can be determined by phase angles dissimilarity at the buses near to the generator. The frequencies that represent the dynamic response of every generator after grid disturbances can be defined as:

$$s_{i,j} = \int_0^T \bigl(\Delta\theta_i(t) - \Delta\theta_j(t)\bigr)\,\mathrm{d}t \tag{1}$$

where $\theta_i$ and $\theta_j$ denote the phase angles at bus $i$ and bus $j$, respectively; $T$ is the observation time; and $s_{i,j}$ is the dissimilarity index between bus $i$ and bus $j$.

The amount of energy observed or delivered by generators in the power system can be reflected by their speed deviations [20]. Therefore, analyzing these frequencies, which represent the dynamic response of generators following a disturbance, can be helpful for coherency determination. These frequency components can be extracted using the discrete Fourier transform (DFT) as follows:

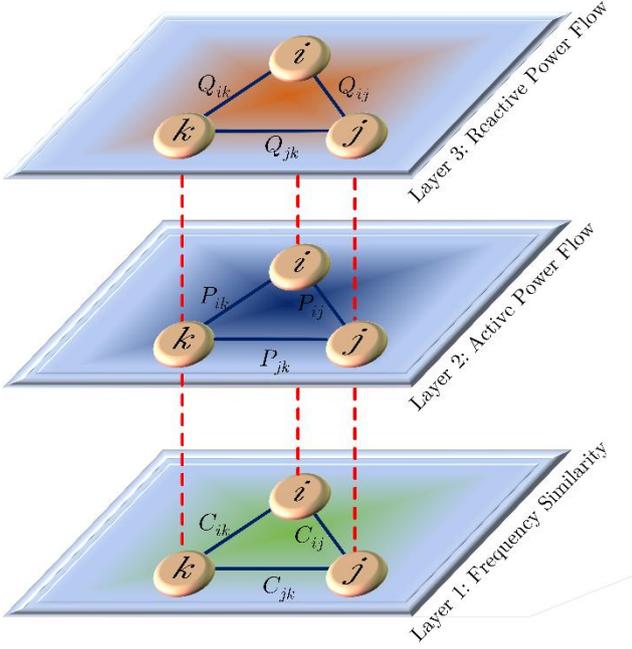

**Fig. 1** Three-layer graph of power networks

$$F_i(f) = \int_0^{N-1} \omega_i(k) e^{-j\frac{2\pi fk}{N}} dk, f = 0,1,\dots,N-1 \quad (2)$$

$$\omega_i(k) = \frac{\theta_i(k)-\theta_i(k-1)}{\Delta t} \quad (3)$$

where $\omega_i(k)$ is the angular velocity of generator $i$ at time instant $k$; $F_i(f)$ is the Fourier transform of the angular speed; $N$ is the number of samples in the waveform; and $\Delta t$ is the time interval between two consecutive samples, held constant throughout simulations.

The vector-space $\boldsymbol{F}_i = [F_i(1), F_i(2), \dots, F_i(N)]^\mathrm{T}$ and $N_B \times N$ dimension matrix $\boldsymbol{F}$ are formed as:

$$\boldsymbol{F} = [\boldsymbol{F}_1, \dots, \boldsymbol{F}_i, \dots, \boldsymbol{F}_{N_B}]^\mathrm{T} \quad (4)$$

where $N_B$ is the total number of buses in the power grid.

The phase angle and the amplitude of each frequency component in the angular velocity signal can be extracted by using the DFT. Therefore, the correlation of the phase angle oscillation of generator/non-generator buses can reveal the coherency of oscillations related to generators, which will be discussed in the next section.

### 2.3 Correlation coefficient similarity matrix

Pearson product-moment correlation coefficient $C_C$ is a popular metric used to evaluate the strength of the association between two variables [5, 16]. The $C_C$ ranges between $-1.0$ and $+1.0$ and quantifies the direction and strength of the linear association between two multidimensional random variables. In connection with power systems, this factor represents the association between two different electrical buses, as shown in (5). A larger $C_{C,ij}$ indicates a stronger connection or higher coherency between bus $i$ and bus $j$.

$$C_{C,ij} = \frac{\sum_{f=1}^{n}[(F_i(f)-F_{i,avg})(F_j(f)-F_{j,avg})]}{\sqrt{\sum_{f=1}^{n}(F_i(f)-F_{i,avg})^2 \times \sum_{f=1}^{n}(F_j(f)-F_{j,avg})^2}} \quad (5)$$

where $C_{C,ij}$ is the correlation coefficient among buses $i$ and $j$; $n$ is the number of frequency components comprised in range; and $F_{i,avg}$ is the average of the frequency components of bus $i$ in the domain of inter-area oscillation modes. We define the correlation coefficient similarity matrix $\boldsymbol{M}_{CCSM}$, as a matrix that its components are equal to $C_{C,ij}$ $i,j \in N_B$, namely,

$$\boldsymbol{M}_{CCSM} = \begin{bmatrix} C_{C,11} & \cdots & C_{C,1N_B} \\ \vdots & \ddots & \vdots \\ C_{C,N_B 1} & \cdots & C_{C,N_B N_B} \end{bmatrix} \quad (6)$$

### 2.4 Reactive power similarity matrix

Generally, in a power system, the voltage and the frequency are controlled by reactive power and active power, respectively. Therefore, considering reactive power and active power simultaneously in the islanding problem would result in more stable islands in terms of frequency and voltage. In order to accomplish the aforementioned goal, the minimal power flow disruption, as shown in (7), can be utilized for controlled islanding as the objective function.

$$\min_{V_1,V_2 \subset V} \left( \sum_{i \in V_1, j \in V_2} |Q_{ij}| \right) \quad (7)$$

where $Q_{ij}$ is the reactive power flow between bus $i$ and bus $j$.

The controlled islanding problem with the above objective function can be transformed into a graph-cut problem by defining a squared $N_B \times N_B$ adjacency matrix with elements $|Q_{ij}|$. Accordingly, a reactive power graph similarity matrix $\boldsymbol{M}_Q$ is defined as:

$$M_{Q,ij} = \begin{cases} \frac{|Q_{ij}|+|Q_{ji}|}{2} = |V_i||V_j||G_{ij}\cos(\phi_i-\phi_j)| & i \neq j \\ 0 & i = j \end{cases} \quad (8)$$

where $V_i$ and $V_j$ are the voltage amplitudes of nodes $i$ and $j$, respectively; $G_{ij}$ is the real part of the admittance matrix; and $\phi_i, \phi_j$ denote the phase angles between the voltage and the current at the respective nodes.

## 2.5 Active power similarity matrix

Similar to the reactive power, the minimal active power flow disruption, as shown in (9), can be defined and utilized for controlled islanding.

$$\min_{V_1, V_2 \subset V} \left( \sum_{i \in V_1, j \in V_2} |P_{ij}| \right) \quad (9)$$

where $P_{ij}$ is the active power flow between bus $i$ and bus $j$.

The controlled islanding problem with the above objective function can be similarly transformed to a graph-cut problem by defining a squared $N_B \times N_B$ adjacency matrix with elements $|P_{ij}|$. Accordingly, an active power graph similarity matrix $M_P$ is defined as:

$$M_{P,ij} = \begin{cases} \frac{|P_{ij}| + |P_{ji}|}{2} = |V_i||V_j||B_{ij} \sin(\phi_i - \phi_j)| & i \neq j \\ 0 & i = j \end{cases} \quad (10)$$

where $B_{ij}$ is the imaginary part of the network admittance matrix.

Utilizing the minimal power flow disruption as the objective function minimizes the amount of load that must be shed following system splitting. The three proposed similarity matrices $M_{CCSM}, M_Q$ and $M_P$, are calculated based on real-time power system data. The use of the aforementioned similarity matrices in one model, one can be anticipated that an appropriate combination of information included in the multiple graph layers would lead to an improved clustering, i.e., this will lead to more precise predictions on the location and extension of the island of stability.

## 3 Controlled islanding via multi-layer spectral clustering while addressing generator's coherency

### 3.1 Stage I: coherency detection based on modularity clustering

Based on the concept of tight coherency, the phase angles of all buses in an area should have relatively the same deviation. This can be assessed by calculating the correlation between each pair of buses in the area using (5). To identify coherency of buses, it is necessary to find strongly connected groups of buses since groups that are strongly coupled tend to maintain synchronism. Online coherency detection based on modularity clustering algorithm will be used to achieve this purpose. It neither requires a predefined number of groups nor a defining threshold value. The objective of this method is to separate the network into groups of vertices that have weak connections between them and to look for the naturally occurring groups in a network regardless of the number size. Greedy optimization of modularity tends to form very fast clustering.

The modularity is defined as the number of edges falling within groups minus the expected number in an equivalent network with edges placed at random. The modularity, denoted by $Q$, is given by:

$$Q = \frac{1}{2m} \sum_{ij} \left[ w_{ij} - \frac{d_i d_j}{2m} \right] \delta(C_i, C_j) \quad (11)$$

where $w_{ij}$ is the weight of the edge between $i$ and $j$; $d_i$ and $d_j$ are the degrees of the vertices $i$ and $j$, respectively; $m$ is the total number of the edges; and $\delta$-function is 1 if nodes $i$ and $j$ are in the same community ($C_i = C_j$), otherwise, it is 0. The value of $Q$ lies in the range [-1,1]. The cluster structure can be searched precisely by checking the network divisions that have large modularity values.

The first step in evaluating coherency of buses of a power network at any point in time is to calculate the correlation coefficient among all the buses and form the correlation coefficient similarity matrix. Then, $k$ groups of coherent buses can be achieved by applying modularity clustering on the correlation coefficient similarity matrix.

### 3.2 Stage II: controlled islanding while preserving coherent bus groups

In graph theory, spectral clustering treats the data clustering as a graph partitioning problem, which is equivalent to minimizing weights of graph cuts. Further, the normalized cuts algorithm can be used to find the solution to the normalized cuts problem. It substantially corresponds to working with the eigenvectors and eigenvalues of the normalized graph Laplacian. The normalized graph Laplacian matrix $L$ is of broad interests in the studies of spectral graph theory and is defined as:

$$\boldsymbol{L} = \boldsymbol{D}^{\frac{1}{2}}(\boldsymbol{D} - \boldsymbol{W})\boldsymbol{D}^{-\frac{1}{2}} \quad (12)$$

where $\boldsymbol{D}$ is the degree matrix, i.e., a diagonal matrix with the vertex degrees along the diagonal that are defined as $D_{ij} = \sum_{j=1}^{M} A_{ij}$, $A_{ij}$ is the component of the adjacency matrix $\boldsymbol{A}$ of $G$; and $\boldsymbol{W}$ is the adjacency matrix.

We consider now the problem of clustering $N_B$ vertices, $V = \{v_i, i = 1,2, \ldots, N_B\}$ of $G$ into $k$ distinct subsets such that the bus nodes in the same subset are similar, i.e., they are connected by edges of large weights. Reference [21] proved that all normalized Laplacian eigenvalues of a graph lie in the interval [0, 2], and 0 is always a normalized Laplacian eigenvalue, a property favorable in comparing different graph layers. We note that the spectral clustering algorithms can efficiently solve this problem. Precisely, we concentrate on the algorithm suggested in [19], which

solves the following trace minimization problem:

$$\min_{U \in \mathbf{R}^{N_B \times k}} tr(U^T L\ U), \quad s.t. \quad U^T U = I \quad (13)$$

where $k$ is the target number of clusters, and $N_B$ is the total number of vertices in the graph.

The clustering of the vertices in $G$ is then implemented using the $k$-means clustering algorithm to the normalized row vectors of the matrix $U$.

Given a multi-layer graph $G$ with $M$ individual layers $\{G_i, i = 1,2, \ldots M\}$, we first compute the graph Laplacian matrix $L_i$ for each $G_i$ and then represent each $G_i$ by the spectral embedding matrix $U_i \in \mathbf{R}^{N_B \times k}$ from the first $k$ eigenvectors of $L_i$.

The goal is to merge these multiple subspaces in a meaningful and efficient way. To merge these multiple subspaces, the Riemannian squared projection distance between the target representative subspace $U$ and the $M$ individual subspaces $\{U_i, i = 1,2, \ldots, M\}$ is computed as the sum of the squared projection distances between $U$ and each individual subspace given by $U_i$:

$$d_{proj}^2 = \sum_{i=1}^{M}[k - tr(UU^T U_i U_i^T)]$$
$$= kM - \sum_{i=1}^{M}[(UU^T U_i U_i^T)] \quad (14)$$

By solving the following optimization problem that integrates both (13) and (14), multiple subspaces can be merged. This method is based on the following Rayleigh-Ritz theorem, which transforms the generalized eigenvalues problem into a constrained minimization problem, described as:

$$\min_{U \in \mathbf{R}^{N_B \times k}} tr\left[U^T \left(\sum_{i=1}^{M} L_i - \alpha \sum_{i=1}^{M} U_i U_i^T\right) U\right] s.t. UU^T = I \quad (15)$$

where $\alpha$ is the weighting parameter that balances the trade-off linking the two terms in the objective function. We may note that this is identical trace minimization problem as introduced in (13), but with a "modified" Laplacian given as:

$$L_m = \sum_{i=1}^{M} L_i - \alpha \sum_{i=1}^{M} U_i U_i^T \quad (16)$$

The proposed M-SCCI algorithm is described as follows.
1) First stage
Step 1: formulate the multi-layer graph $G$ using only bus nodes, with edge weights equal to the $C_{C,ij}$, $M_{P,ij}$ and $M_{Q,ij}$.
Step 2: obtain the $k$ cluster groups of coherent buses from Step 1.
2) Second stage
Step 3: input $N_B \times N_B$ weighted adjacency matrices $\{W_i, i = 1,2, \ldots, M\}$ of each individual graph layers $\{G_i, i = 1,2, \ldots, M\}$, $k$, and $\alpha$.
Step 4: calculate the normalized Laplacian matrix $L_i$ and the subspace illustration $U_i$ for each $G_i$.
Step 5: compute the graph Laplacian matrix $L_m$ with (16).
Step 6: compute $U \in \mathbf{R}^{N_B \times k}$.
Step 7: normalize each row of $U$ to get $U_{norm}$.
Step 8: let $y_j \in \mathbf{R}^k$ ($j = 1,2, \ldots, n$) be the transpose of the $j$-th row of $U_{norm}$.
Step 9: cluster $y_j$ into $C_1, C_2, \ldots, C_k$ using the $k$-means algorithm.
Step 10: output cluster assignments $C_1, C_2, \ldots, C_k$.

This algorithm uses the correlation coefficient between the frequency components among $C_{C,ij}$, $M_{P,ij}$, and $M_{Q,ij}$ data to produce an islanding solution with minimal power flow disruption. In the first stage, the buses are grouped using modularity clustering, based on the $C_{C,ij}$. The number of $k$ clusters outcomes of this grouping serve as the input to the second stage, in which nodes are grouped based on multi-layer constrained spectral clustering. The M-SCCI algorithm proposed here can identify, in real time, an islanding solution that has minimal power flow disruption and satisfies the bus coherency constraints.

## 4 Simulation studies

The model effectiveness is evaluated through the simulation study conducted on the modified IEEE 39-bus system. The methodology has been implemented in MATLAB and all time-domain simulations are achieved in DIgSILENT PowerFactory. To stress the system and raise the likelihood of instability following a disturbance, we increased the base load level by 25% at 0.01 s. Then, two short circuit events occurred in lines 13-14 and 16-17 at 2 s. The short circuit events are cleared after 0.20 s by opening the line switches from the substations, while the simulation lasts for 5 s.

Figures 2 and 3 demonstrate the rotor angle of generator and the system frequency, respectively, that indicate the system instability following the short circuit events. The proposed solution approach is applied to the system to determine the islanding boundaries. The quality of each island is then evaluated by calculating the dynamic behavior and the power mismatch in the islands. It can be observed from Fig. 3, that if no control action is undertaken, the system loses synchronism at about 2.25 s. Indeed, real-time simulation in DIgSILENT indicates out of step at 2.25 s for generators. As noticed, the system is divided into two groups, which are not balanced.

The frequency of the generators and the loss of synchronism are a clear indication that the system should be split.

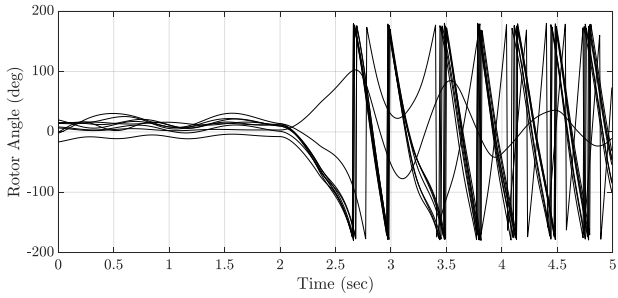

**Fig. 2** Rotor angle after two short circuit events without islanding

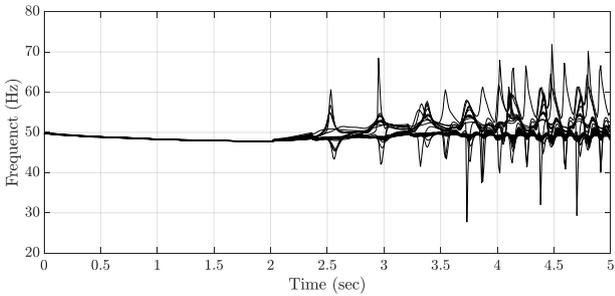

**Fig. 3** System frequency after two short circuit events without islanding

Before proceeding to discuss these case studies for our proposed methodology, we examine the islanding methodology proposed in [9] to split the network. In this method, the authors proposed a two-step constrained spectral clustering-controlled islanding to find the islanding solution, which provided the minimum power flow disruption while satisfying the constraint of coherent generator groups.

As it will be shown, in the following example, a multiple variant of valid cut-sets separating coherent generator groups from each other is possible, but only certain variants will allow secure islanding.

According to the proposed model in [9], it is essential to find the minimum cut in a graph that its edges are the active power distortion and constraint the clusters with the coherent groups of generators. Accordingly, we need to first find the coherent groups of generators and then establish the connectivity constraints among the generators within a group and non-connectivity constraints between the generators in different groups. Finally, the spectral constraint clustering is applied to the problem and determine the islands in the power system.

Following clearing the fault and applying the modularity clustering to the $K_s$ matrix proposed in [16] at 2.21 s, two coherent groups of generators, {G1, G2, G3, G8, G9, G10} and {G4, G5, G6, G7}, are produced. The two coherent groups of generators form the set of connectivity and non-connectivity constraints, in which all pairs of generators in one group must be linked together (connectivity constraints), and generators in different groups must not be linked with each other (non-connectivity constraints), where the associated schematic is shown in Fig.4.

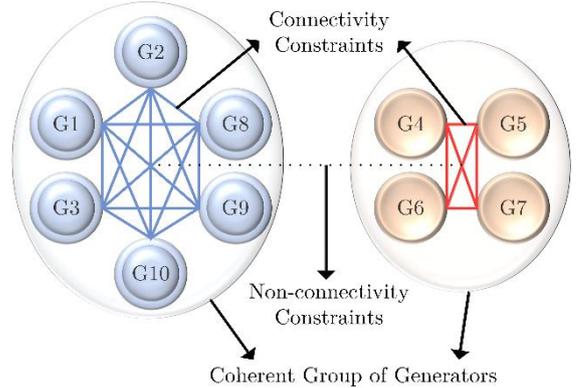

**Fig. 4** Coherent groups of generators and constraints of connectivity and non-connectivity

After determining the coherent groups of generators, the active power graph similarity matrix is clustered into two groups, using the constraint clustering approach [22], where the clusters outcome and the rotor angles of generators following the clustering are depicted in Fig. 5, and Fig. 6, respectively.

As can be seen, while the generators in island 2 are stable, generators in island 1 become out of steps. On the other hand, the graph-based islanding solutions automatically satisfy the connectivity constraints, since the solution is sought through the minimum cuts in the graph. Our approach is a graph-based approach, wherein in each island the nodes preserve their pre-islanding conditions, and the network is separated by cutting the edges in the graph.

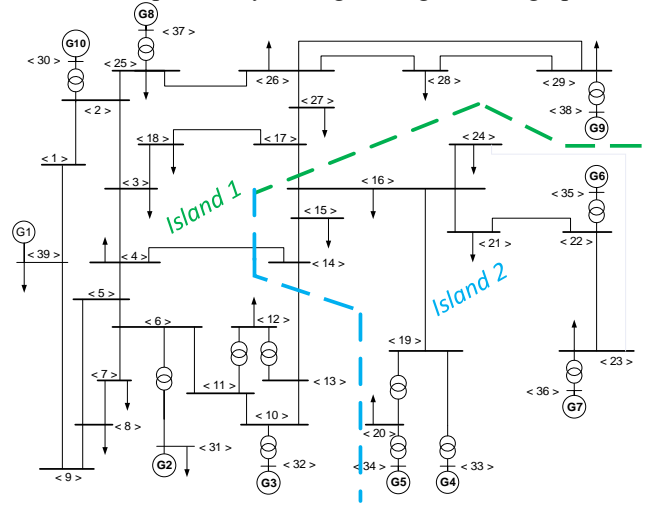

**Fig. 5** Islanding boundaries after applying the clustering method

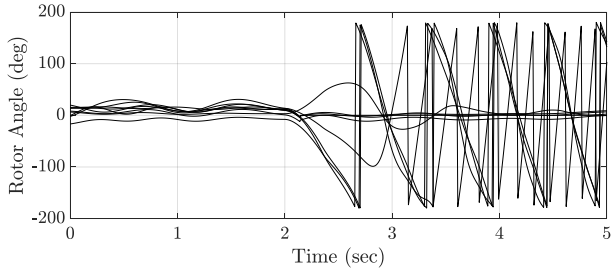

**Fig. 6** Rotor angles of generators after applying the clustering method

The following sub-sections compare the result using three different criteria, i.e. frequency similarity, reactive power, and active power, for the islanding decision making procedure. Four cases under the same operating conditions are employed to demonstrate the accuracy of the proposed M-SCCI algorithm.

Case 1: single-layer intentional islanding based on frequency similarity of the island's buses.

Case 2: single-layer intentional islanding based on reactive power.

Case 3: single-layer intentional islanding based on active power.

Case 4: multi-layer intentional islanding based on all three criteria.

In all cases, the islanding scheme is applied at t=2.21 sec, just after clearing the fault to avoid generator instability that happens at t=2.25 sec if no action is taken.

### 4.1 Case 1

In this case study, the frequency similarity is employed as the main criterion for islanding decision making. The approach provides a suitable islanding solution using online coherency and pre-fault power flow conditions. In the first stage, the proposed buses coherency modularity clustering algorithm based on frequency similarity of the island's buses identified two sets of coherent generators {G1, G2, G3, G10} and {G4, G5, G6, G7, G8, G9}.

Considering these two sets found in first stage, the number of two clusters outcomes serves as the input in the second stage to solve the single-layer constrained spectral clustering. The islanding solution suggests that it should be split into two islands as shown in Fig. 7. The resulted groups using $C_{C,ij}$ are two groups as depicted in Fig. 8 with two background colors.

The allocation of buses in Case 1 to coherent generator groups is as follows: ① island 1, buses B1-B14, B30-B32, and B39; ② island 2, buses B15-B29 and B33-B38. The minimal power flow disruption across boundaries of islands are 857 MW active power and 1349 Mvar reactive power.

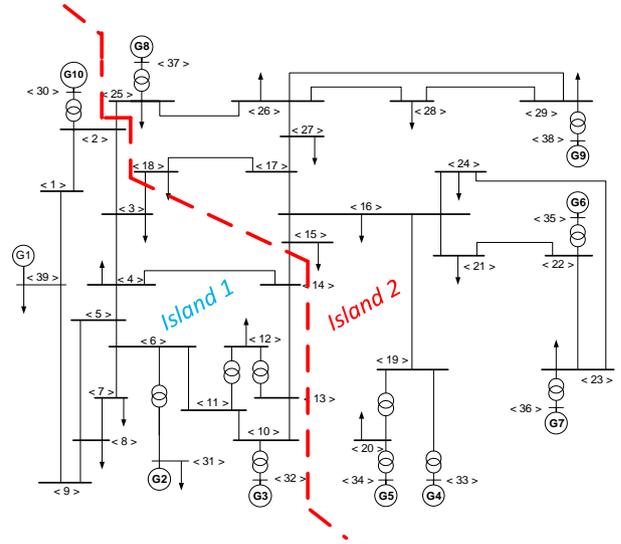

**Fig. 7** Islanding boundaries considering frequency similarity

Figure 9 shows the generator rotor angle oscillation during the simulation study of 15 s. Obviously, the rotor angle oscillations are unstable, and all the machines lose synchronism while groups of generators become weaker following the events.

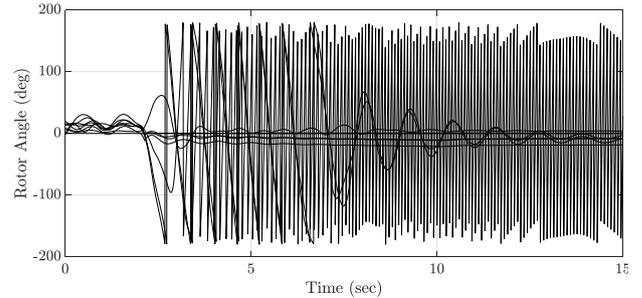

**Fig. 8** Rotor angles after islanding based on frequency similarity

### 4.2 Case 2

In this case study, the minimum reactive power flow disruption is employed as the main criterion for islanding decision making procedure. The system initial condition is the same as that of Case 1. The same faults as that of Case 1 are imposed. In the first stage of the proposed bus coherency, modularity clustering algorithm based on reactive power similarity of the island's buses identified three sets of coherent generators {G1, G2, G3, G10}, {G4, G5, G6, G7}, and {G8, G9}.

Considering these three sets found in the first stage, the number of three clusters outcomes serves as the input in the second stage to solve the single-layer constrained spectral clustering based on the minimum reactive power flow disruption. The islanding solution suggests that it should be three islands as shown in Fig. 10. The resulted groups using $M_{Q,ij}$ are three groups as depicted in Fig. 11 with three background colors.

**Fig. 9** Correlation coefficient similarity matrix at 2.21 s

**Fig. 10** Islanding boundaries considering reactive power

The allocation of buses in Case 2 to coherent generator groups is as follows: ① island 1, buses B1-B15, B18, B30-B32, and B39; ② island 2, buses B16, B19-B24, and B33-B36; ③ island 3, buses B17, B25-B29, B37, and B38. The minimal power flow disruption across boundaries of islands are 2291 MW active power and 1349 Mvar reactive power.

Figure 12 shows the generator rotor angle oscillation during the simulation study of 15 s. Obviously; the rotor angle oscillations are damped, and all the machines lose synchronism while groups of generators became weaker following the events.

**Fig. 11** Rotor angles after islanding based on reactive power

### 4.3 Case 3

In this case study, the minimum active power flow disruption is employed as the main criterion for islanding decision making procedure. The system initial condition and the fault are the same as that of Case 1.

The first stage of the proposed bus coherency modularity clustering algorithm based on active power similarity of the island's buses returned three coherent

**Fig. 12** Reactive power similarity matrix at 2.21 s

generator groups {G1, G2, G3, G8, G10}, {G4, G5, G6, G7}, and {G9}.

**Fig. 13** Islanding boundaries considering active power

Considering these three sets found in the first stage, the number of 3 clusters outcomes serves as the input in the second stage to solve the single-layer constrained spectral clustering based on the minimum active power flow disruption. The resulted groups, using the $M_{P,ij}$ are three groups as depicted in Fig. 13 with three background colors. The islanding solution suggests that there should be three islands as shown in Fig. 14.

The allocation of buses in Case 3 to coherent generator groups is as follows: ① island 1, buses B1-B15, B25, B30-B32, and B39; ② island 2, buses B16-B24 and B33-B36; ③ island 3, buses B26-B29 and B38. The minimal power flow disruption across boundaries of islands are 1108 MW active power and 1349 MVar reactive power are disrupted.

Figure 15 shows the generator rotor angle oscillation during the simulation study of 15 s. Obviously, the rotor angle oscillations are damped, and all the machines lose synchronism. As can be seen, the power system is not stable.

**Fig. 14** Rotor angles after islanding based on active power

Fig. 15 Active power similarity matrix at 2.21 s

### 4.4 Case 4

The system initial condition and the fault are the same as all the previous cases. In this case study, the frequency, active power, and reactive power similarity matrices are employed as the main criteria for islanding decision making procedure. We implemented intentional islanding at 2.21 s following two cascading outages. First, correlation coefficient is calculated using (5) for all pairs of buses which result in the $C_{C,ij}$ shown in Fig. 8. Applying the modularity clustering on the $C_{C,ij}$ returned two coherent generator groups {G1, G2, G3, G8, G9, G10} and {G4, G5, G6, G7}.

Considering these two sets found in the first stage, the number of two clusters outcomes serves as the input in the second stage to solve the M-SCCI. Then, in the second stage, the M-SCCI algorithm was excused using the three-layer graph with weighted adjacency matrices, i.e. $M_{CCSM}$, $M_Q$ and $M_P$, as the main criteria for islanding decision making procedure, taking into consideration the two cluster coherency groups found in the first stage of the algorithm. The allocation of buses in Case 4 to the coherent generator groups is as follows: ① island 1, buses B1-B14, B17, B18, B25-B32, and B37-B39; ② island 2, buses B15, B16, B19-B24, and B33-B36.

Fig. 16 Islanding boundaries based on multi-layer clustering

The final splitting strategy, possessing the lowest power exchange is represented in Fig. 16. The minimal power flow disruption across boundaries of islands are 622 MW active power and 1349 Mvar reactive power. Figure 17 shows the generator rotor angle oscillation during the simulation study of 15 s. Obviously, the rotor angle oscillations are damped, and all the machines remain in synchronism while groups of generators became stronger following the events. Table 1 presents the power flow mismatch between the islands for each case study.

The comparison in Table 1 shows that the proposed M-SCCI algorithm using all criteria returns the cut-set that separated the coherent generator groups with minimum cut, which is 622 MW.

Table 1 Summary of power flow mismatch between islands for each case study

| Clustering criteria | Island | Active power (generator) $P_G$ (MW) | Active power (load) $P_L$ (MW) | Reactive power (generator) $Q_G$ (MVar) | Reactive power (load) $Q_L$ (MVar) | ΔP (MW) | ΔQ (MVar) | $\sum|\Delta P|$ (MW) | $\sum|\Delta Q|$ (Mvar) |
|---|---|---|---|---|---|---|---|---|---|
| Clustering based on reactive power | 1 | 3672 | 4765 | 1663 | 1458 | -1092 | 205 | 2291 | 1349 |
| | 2 | 3525 | 2681 | 1397 | 364 | 844 | 1032 | | |
| | 3 | 2055 | 1700 | 402 | 291 | 355 | 111 | | |
| Clustering based on active power | 1 | 4482 | 4864 | 1802 | 1484 | -381 | 318 | 1107 | 1349 |
| | 2 | 3525 | 2918 | 1397 | 409 | 607 | 987 | | |
| | 3 | 1245 | 1364 | 264 | 221 | -119 | 44 | | |
| Clustering based on frequency similarity | 1 | 5580 | 5098 | 1799 | 930 | 482 | 869 | 857 | 1349 |
| | 2 | 3672 | 4048 | 1663 | 1183 | -375 | 480 | | |
| Clustering based on all three criteria, $k$=3 | 1 | 2055 | 1700 | 402 | 291 | 355 | 111 | 1331 | 1349 |
| | 2 | 3525 | 3161 | 1397 | 594 | 364 | 803 | | |
| | 3 | 3672 | 4285 | 1663 | 1228 | -612 | 435 | | |
| Clustering based on all three criteria, $k$=2 | 1 | 5727 | 5985 | 2066 | 1520 | -258 | 546 | 622 | 1349 |
| | 2 | 3525 | 3161 | 1397 | 594 | 364 | 803 | | |

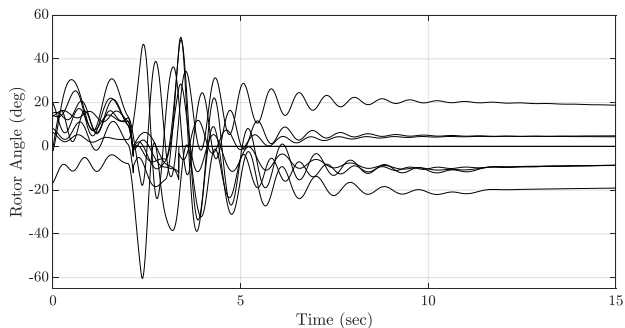

Fig. 17 Rotor angles after islanding based on all criteria

## 5 Conclusion

This paper proposed a computationally efficient real-time ICI algorithm based on multi-layer graphs, subspace analysis, and constrained spectral clustering while addressing the generator coherency problem. We demonstrated that using multi-layer spectral clustering to find the islanding boundaries, instead of using a single layer, i.e., the frequency similarity, the active power, and the reactive power produced improved clustering performance. The insertion of the bus coherency constraints prevents new island groupings that would contain non-coherent generators. The use of minimal power-flow disruption improves the transient stability of the islands produced. The simulation results show that the proposed M-SCCI algorithm is computationally efficient and is suitable for use in real-time applications involving large power systems.